\documentclass[12pt]{article}

\usepackage{scicite}
\usepackage{graphicx}
\usepackage{times}

\newenvironment{sciabstract}{%
\begin{quote} \bf}
{\end{quote}}

\renewcommand\refname{References and Notes}
\newcommand{\degree}{$^\circ$~}
\newcounter{lastnote}

\title{Imaging the Surface of Altair} 

\author
{John D. Monnier$^{1\ast}$,  M. Zhao$^{1}$, E. Pedretti$^{2}$, N. Thureau$^3$, M. Ireland$^4$,\\
P. Muirhead$^5$, J.-P. Berger$^6$, 
R. Millan-Gabet$^7$,\\  G. Van Belle$^7$, T. ten Brummelaar$^8$, H. McAlister$^8$, 
S. Ridgway$^9$, \\N. Turner$^8$, L. Sturmann$^8$, J. Sturmann$^8$, 
D. Berger$^{1}$\\
\\
\normalsize{$^{1}$Department of Astronomy, University of Michigan,}\\
\normalsize{$^{2}$University of St. Andrews, Scotland, UK}\\
\normalsize{$^{3}$University of Cambridge, Cambridge, UK}\\
\normalsize{$^{4}$California Institute of Technology, Pasadena, CA}\\
\normalsize{$^{5}$Cornell University, Ithaca, NY}\\
\normalsize{$^{6}$Laboratoire d'Astrophysique de Grenoble, France}\\
\normalsize{$^{7}$Michelson Science Center, Pasadena, CA}\\
\normalsize{$^8$CHARA, Georgia State University, Atlanta, GA}\\
\normalsize{$^{9}$National Optical Astronomy Observatory, Tucson, AZ}\\
\normalsize{$^\ast$To whom correspondence should be addressed; E-mail:  monnier@umich.edu.}
}

\date{}

\begin{document} 

% Double-space the manuscript.Another Unknown Address, Palookaville, ST 99999, USA

\baselineskip24pt


\maketitle


\begin{sciabstract}
  Spatially resolving the surfaces of nearby stars promises to
  advance our knowledge of stellar physics.  Using optical long-baseline 
  interferometry, we present here a
  near-infrared image of the rapidly rotating hot star Altair with
  $<$1 milliarcsecond resolution.  The image clearly reveals the strong effect
  of gravity darkening on the highly-distorted stellar photosphere.
  Standard models for a uniformly rotating star can not explain our results, 
  requiring 
differential rotation, alternative gravity darkening laws, or both.

\end{sciabstract}

% In setting up this template for *Science* papers, we've used both
% the \section* command and the \paragraph* command for topical
% divisions.  Which you use will of course depend on the type of paper
% you're writing.  Review Articles tend to have displayed headings, for
% which \section* is more appropriate; Research Articles, when they have
% formal topical divisions at all, tend to signal them with bold text
% that runs into the paragraph, for which \paragraph* is the right
% choice.  Either way, use the asterisk (*) modifier, as shown, to

While solar astronomers can take advantage of high-resolution,
multi-wavelength, real-time imaging of the Sun's surface, stellar
astronomers know most stars, located parsecs or kiloparsecs away, as
simple points of light.  To discover and understand the novel
processes around stars unlike the Sun, we must rely on stellar spectra
averaged over the entire photosphere. Despite their enormous value,
spectra alone have been inadequate to resolve central questions in
stellar astronomy, such as the role of angular momentum in stellar
evolution\cite{MM2000}, the production and maintenance of magnetic
fields\cite{landstreet1992},
the launching of massive stellar winds\cite{kervella2006}, and the
interactions between very close binary companions\cite{barai2004}.

Fortunately, solar astronomers no longer hold a monopoly on stellar
imaging.  Using long-baseline visible and infrared interferometers,
the photospheric diameters of hundreds of stars and high precision  
dynamical masses for dozens of binaries have been catalogued, offering
exacting constraints for theories of stellar evolution and stellar
atmospheres\cite{monnier2003}.  This work requires an angular
resolution of $\sim$1 milliarcseconds (1 part in 2$\times$10$^8$, or 5
nano-radians) for resolving even nearby stars and is more than
an order-of-magnitude better than that achievable with the Hubble Space
Telescope or ground-based 8-m class telescopes equipped with adaptive
optics. 

Stellar imaging can be used to investigate rapid rotation of hot, massive 
stars.
A significant fraction of hot stars are rapid rotators with
surface rotational velocities of more than 100 km/s
\cite{abt1995,abt2002}.  These rapid rotators are expected to traverse
very different evolutionary paths than their slowly rotating
kin\cite{MM2000} and rotation-induced mixing alters 
stellar abundances\cite{pinsonneault1997}.  While hot stars are
relatively rare by number in the Milky Way Galaxy, they have a disproportional
effect on galactic evolution due to their high luminosities, strong
winds, and their final end as supernovae (for the most massive stars).
Recently, rapid rotation in single stars has been invoked to explain at
least one major type of gamma ray bursts\cite{brown2000} and binary
coalescence of massive stars/remnants for another\cite{gehrels2005}.

The distinctive observational signatures of rapid rotation were first described
by 
von Zeipel \cite{vz1924b}, beginning with the expection that centrifugal forces
would distort the photospheric shape
and that the resulting oblateness
would induce lower effective temperatures at the equator.  This latter
effect, known as gravity darkening, will cause distortions in the observed
line profiles as well as the overall spectral energy
distribution.
Precise
predictions can be made but rely on uncertain assumptions, most critically the
distribution of angular momentum in the star -- uniform rotation is often
assumed  for simplicity.

The most basic predictions of von Zeipel theory -- centrifugal
distortion and gravity darkening -- have been confirmed to some
extent.  
The Palomar Testbed Interferometer (PTI) was first to measure
photospheric elongation in a rapid rotator, finding 
the diameter of the nearby A-type star
Altair to be $\sim$14\% larger in one dimension than
the other\cite{vanbelle2001}. 
The Navy Prototype Optical
Interferometer (NPOI) and the Center For High Angular Resolution
Astronomy (CHARA) interferometric array both measured strong limb-darkening 
profiles for the photometric standard
Vega \cite{peterson2006a,aufdenberg2006}, 
consistent with rapid rotator viewed nearly pole-on.
A brightness asymmetry for Altair was also reported by NPOI 
\cite{peterson2006b,ohishi2004}, suggestive of the expected
pole-to-equator temperature difference from gravity darkening.
In recent years, a total of five
rapid rotators have been measured to be elongated by interferometers
\cite{souza2003,mcalister2005,vanbelle2006}.

While von Zeipel theory appears to work at a basic level, serious discrepancies
between theory and observations have emerged.
Most notably, the diameter of the B3V-type star Achernar
\cite{souza2003} was measured to be $\sim$1.56 times longer in one dimension
than the other, too large to
be explained by von Zeipel theory.  Explanations for this include
strong differential rotation of the star\cite{jackson2004} or the
presence of a polar wind\cite{kervella2006}, either of which have
far-reaching consequences for our understanding of stellar evolution.
In order to address these issues, we must move beyond the simplest 
models for rapidly rotating stars, and this will require a corresponding
jump in the quality and quantity of interferometry data.
Indeed, all previous results were based on 
limited interferometer baselines, lacking the capability to form
model-independent images, relying entirely on model-fitting for interpretation.
Thus previous confirmations
of von Zeipel theory, although suggestive, were incomplete.

Here we report a development in imaging capabilities that tests the
von Zeipel theory, both through basic imaging and precise
model-fitting.  By combining near-infrared light from four telescopes
of the Georgia State University CHARA interferometric array, we have
synthesized an elliptical aperture with dimensions 265x195 meters
(Figure 1), allowing us to reconstruct images of the prototypical
rapid rotator Altair (spectral type A7V) with an angular resolution of
$\sim$0.64 milliarcseconds, the diffraction limit defined by $\frac{\lambda}{2D}$,
the observing wavelength divided by twice the longest interferometer baseline.
The recently-commissioned Michigan Infrared Combiner (MIRC)
\cite{mirc2004} was essential for this work, 
allowing the light from the CHARA telescopes to be all combined together
simultaneously in 8 spectral channels spanning the astronomical H band
($\lambda=1.50-1.74\mu$m).  
The Altair data presented here were 
collected on UT2006Aug31 and
UT2006Sep01 -- complete observational information is available \cite{altair_som}.
In addition, we utilized some K band ($\lambda=2.2\mu$m)
observations by the PTI to constrain the
short-baseline visibilities in subsequent analysis.  

By using four CHARA telescopes, interferometric imaging of Altair is now possible, 
although
this requires specialized image reconstruction techniques.  We utilized the 
publicly-available application ``Markov-Chain Imager for Optical
Interferometry (MACIM)'' \cite{macim2006} in this work, applying the
Maximum Entropy Method (MEM) \cite{mem1986}.  
We restricted 
the stellar image to fall within an elliptical boundary,
similar in principle to limiting the
field-of-view in standard aperture synthesis procedures.
This restriction  biases our imaging against 
faint emission features arising outside the
photosphere; however,  we do not expect any circumstellar emission in Altair
which is relatively cool, lacking signs of gas emission or strong winds.
Further details of our imaging procedures, along with results from
validation tests, can be found in \cite{altair_som}.

Our image shows the stellar photosphere of Altair to be well-resolved (Figure~2A),
appearing elongated in the NE-SW direction with a bright dominant
feature covering the northwest quadrant of the star.  In order to
reduce the influence of possible low-level artifacts that are beyond
the diffraction-limit of our interferometer, we have followed the
standard procedure \cite{hogbom1974} of convolving the reconstructed
image with a Gaussian beam matching the resolution of the
interferometer (Figure 2B).  

These images confirm the basic picture of gravity darkening induced by
rapid rotation.  We see Altair's photosphere to be oblate with a
bright region identifiable as the stellar polar region.  The intensity
of the dark equatorial band is approximately 60-70\% of the brightness
at the pole, broadly consistent with expectations for the
near-infrared from previous models.  While we see some evidence for
deviations from axisymmetry (small excess emission on northern limb), this 
feature is at the limit of our image fidelity
and will require additional Fourier coverage to investigate further.

We have also fitted our new extensive dataset with a rapid rotator model,
following the prescription set out in Aufdenberg et
al. \cite{aufdenberg2006} and references therein, assuming a Roche
potential (central point mass) and solid body rotation.  The main
parameters of the model are the stellar radius and temperature at the
pole, the angular rotation rate as a fraction of breakup
($\omega$), the gravity darkening coefficient ($\beta$) and the
viewing angles (inclination and position angle). We employed the
stellar atmosphere models of Kurucz \cite{kurucz1993} for determining
the specific intensity of each point on the surface as a function of
local gravity, effective temperature, 
and limb darkening.  In
addition to matching the new MIRC/CHARA data, we forced the model to
match the measured V and H band photometric magnitudes
(0.765$\pm$0.015 and 0.235$\pm$0.043 respectively) derived from a broad
literature survey.
When fixing the gravity-darkening cofficient to $\beta=0.25$ appropriate for
radiative envelopes,  our derived parameters (Table~1) agree well with
best-fit parameters of Peterson et al. \cite{peterson2006b} 
based on visible data. However, our
best-fit model reached only a reduced $\chi^2_\nu$ of 1.79, suggesting a
need for additional degrees of freedom in our model.

In order to improve our fits, we explored an extension to the von Zeipel model, allowing the
gravity darkening coefficient $\beta$ to be a free parameter.  
We found that $\beta=0.190$ model significantly improved the goodness-of-fit
(Table 1) and this improvement is visually apparent when comparing synthetic
model images to the Altair image from CHARA (Figure 3).
In addition to a lower $\beta$, the new model prefers a slightly less inclined
orientation, a cooler polar temperature, and a faster rotation rate.

Both our imaging and modeling results point to important deficiencies 
in the currently-popular models for rapid rotators.
Previous workers have also
encountered problems explaining high-resolution interferometry data with
standard prescriptions for rotating stars. In addition to the
Achernar case previously cited,
Peterson et al. \cite{peterson2006b} were unable to find a satisfactory fit 
for Altair assuming a standard
Roche - von Zeipel model ($\chi^2_\nu = 3.8$), consistent with the need for
additional stellar physics.
Recent results for Alderamin \cite{vanbelle2006} also specifically favor models with
smaller $\beta$s, in line with our findings.

While model fitting has revealed
deviations from standard theory, our model-independent imaging
allows new features to be discovered outside current model paradigms.  
The most striking difference between our CHARA image and the synthetic model images
(Figure~3) is that our image shows stronger darkening along the equator,
inconsistent with any von Zeipel-like gravity darkening prescription 
assuming uniform rotation.  

Lower equatorial surface temperatures could naturally arise 
if the equatorial rotation rate was higher than the rest of the star 
(differential rotation), reducing the effective gravity
at the surface \cite{jackson2005}.  
Another possibility is that the cooler equatorial layers could be unstable to 
convection \cite{claret2000,espinosa2007}, invalidating a single
gravity darkening ``law'' applicable to all stellar latitudes.
Other studies \cite{lovekin2006} have found further faults with simple application of
the von Zeipel law due to opacity effects in the surface layers.

While difficult to isolate or untangle these various effects from one another,
nevertheless the new interferometric results and our modeling convincingly
establishes the case for stellar 
physics beyond the standard models used today to
describe rotating stars.  A path forward is clear: differential rotation
will leave both geometric and kinematic signatures different from
opacity or convection-related phenomena. Observers must
be armed with a new generation of
models incorporating these physical processes in order to 
exploit the powerful combination of detailed line profile analysis and
multi-wavelength interferometric imaging now available.

\newpage

\begin{figure}[ht]
\begin{center} 
\includegraphics[angle=90,width=3in,clip]{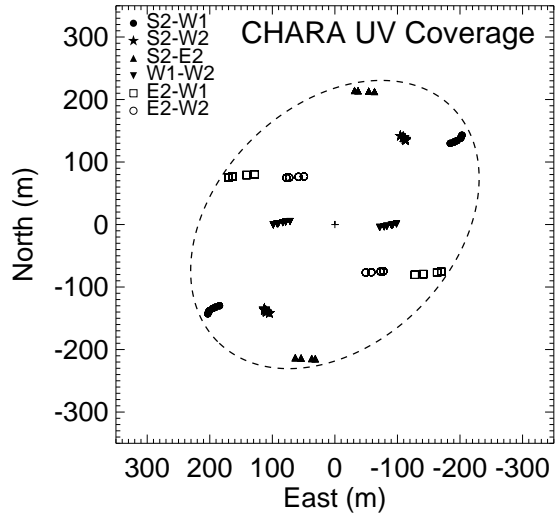}
\caption{\footnotesize This figure shows the Fourier UV coverage
for the Altair observations, where each point represents the projected
separation between one pair of CHARA telescopes (S2-E2-W1-W2) \cite{theo2005}.  
The dashed ellipse shows the
equivalent coverage for an elliptical aperture of 265$\times$195 meters oriented
along a Position Angle of 135\degree East of North.  
\label{fig1}} 
\end{center}
\end{figure}

\newpage

\begin{figure}[ht]
\begin{center}
\includegraphics[angle=90,width=6in,clip]{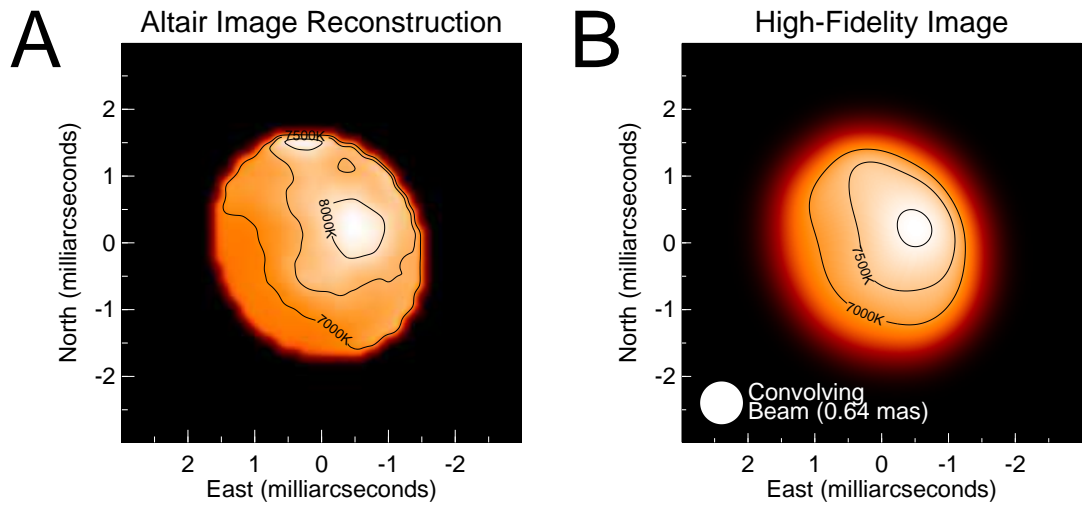}
\caption{\footnotesize A) shows the intensity image of the surface of
Altair ($\lambda=1.65\mu$m)
created with the MACIM/MEM imaging method using a
uniform brightness elliptical prior ($\chi^2_\nu = 0.98$).
Typical photometric errors in the image correspond to $\pm$4\% in intensity.
B) shows the reconstructed image convolved with a
Gaussian beam of 0.64~mas, corresponding to the diffraction-limit of
CHARA for these observations.
For both panels, the specific intensities at 1.65$\mu$m were converted into the corresponding blackbody
temperatures
and contours for 7000K, 7500K, and 8000K are shown. North is up and East is left.
\label{fig2}}
\end{center}
\end{figure}

\newpage

\begin{figure}[ht]
\begin{center}
\includegraphics[angle=90,width=6in,clip]{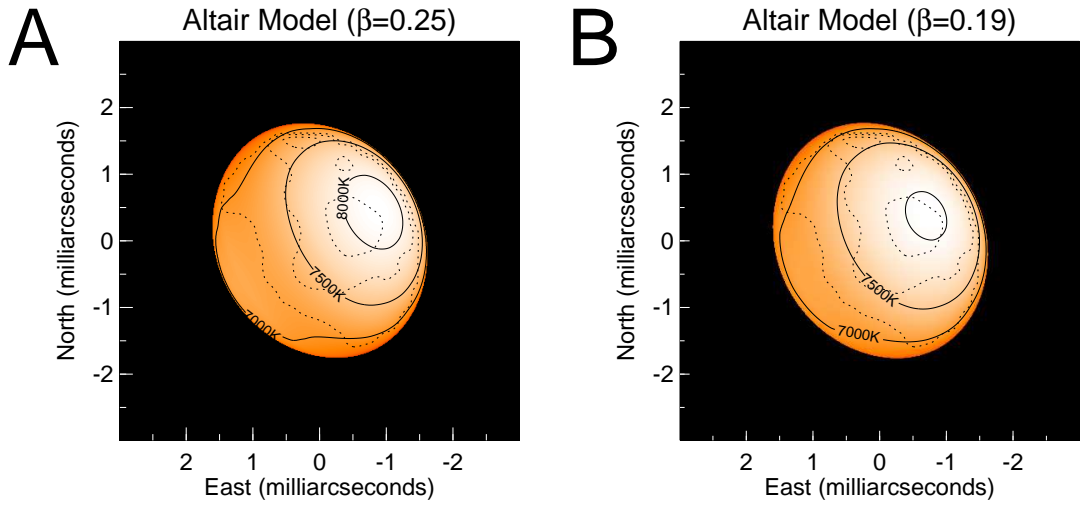}
\caption{\footnotesize The panels show  synthetic images of Altair
($\lambda=1.65\mu$m) adopting conventional rapid rotation models.
A) is the best-fit model assuming standard gravity-darkening
coefficient for radiative envelopes ($\beta=0.25, \chi^2_\nu = 1.79$) while
B) shows the result when $\beta$ is a free parameter
($\beta=0.190, \chi^2_\nu = 1.37$).
For both panels, the specific intensities at 1.65$\mu$m were converted into the corresponding
blackbody temperatures
and contours for 7000K, 7500K, and 8000K are shown.
We have overplotted the contours from the CHARA image (Figure 2A) as dotted lines to facilitiate intercomparison.
\label{fig3}}
\end{center}
\end{figure}

\newpage

\begin{table}

\caption{Best-fit parameters for Roche - von Zeipel models of Altair.  Parameter descriptions: Inclination (0\degree is pole-on, 90\degree is edge-on) \&  Position Angle (degrees East of North)
describe our viewing angle, T$_{\rm pole}$ / R$_{\rm pole}$ describe the temperature and radii of the pole (alternatively, one
can describe the temperature and radii at the equator
T$_{\rm eq}$ / R$_{\rm eq}$ ), $\omega$ is the angular rotation rate as a fraction of critical breakup rate, $\beta$ is the  gravity-darkening coefficient. Models assumed stellar mass 1.791 M$_\odot$ \cite{peterson2006b}, metallicity [Fe/H]$=-0.2$ \cite{erspamer2003}, and distance 5.14 pc \cite{hipparcos}.
}

\begin{center}
\begin{tabular}{lll}
Parameters
 & $\beta$ Fixed & $\beta$ Free \\
\hline
Inclination (degs)    &  62.7 $\pm$ 1.5   & 57.2  $\pm$ 1.9   \\
Position Angle (degs)  & -61.7 $\pm$ 0.9    & -61.8 $\pm$ 0.8   \\
T$_{\rm pole}$ (K)         & 8710   $\pm$ 160    & 8450   $\pm$ 140   \\
R$_{\rm pole}$  (R$_\odot$)   & 1.661  $\pm$ 0.004  &  1.634 $\pm$ 0.011  \\
\qquad         (mas) & 1.503 $\pm$ 0.004   & 1.479 $\pm$ 0.010  \\
T$_{\rm eq}$ (K)         & 6850   $\pm$ 120    & 6860   $\pm$ 150   \\
R$_{\rm eq}$  (R$_\odot$)   & 2.022  $\pm$ 0.009  &  2.029 $\pm$ 0.007  \\
\qquad         (mas) & 1.830  $\pm$ 0.008   & 1.835 $\pm$ 0.007  \\
$\omega$             & 0.902 $\pm$ 0.005   & 0.923 $\pm$ 0.006  \\
$\beta$               & 0.25 (Fixed)       & 0.190 $\pm$ 0.012      \\
Model V Mag      & 0.765               & 0.765 \\
Model H Mag      & 0.225               & 0.220 \\
Model v~$\sin{i}$ (km/s) & 241 & 240 \\
\hline
Reduced $\chi^2$: & & \\
\qquad Total              & 1.79       & 1.37                   \\
\qquad Closure Phase         & 2.08       & 1.73      \\
\qquad Vis$^2$    & 1.48    & 1.10      \\
\qquad Triple Amp   & 2.14       & 1.58        \\
\hline
\end{tabular}
\end{center}
\end{table}

%\hphantom{\cite{ackno}}

\renewcommand\refname{References and Notes}
% The following lines set up an environment for the last note in the
% reference list, which commonly includes acknowledgments of funding,
% help, etc.  It's intended for users of BibTeX or the {thebibliography}
% environment.  Users who are hand-coding their references at the end
% using a list environment such as {enumerate} can simply add another
% item at the end, and it will be numbered automatically.

\clearpage
\newpage
\section*{Supplemental Online Materials}

The Michigan Infrared Combiner (MIRC) is a new instrument on the
Georgia State University CHARA interferometric array; MIRC and the
CHARA Array have both been described in detail
\cite{theo2005,mirc2004} and MIRC commissioning results have
already been presented \cite{mirc2006}.
Here in the
supplemental online materials, we present validation studies of our
data pipeline and image reconstruction methods.

MIRC is an image plane combiner which currently combines light from four
CHARA telescopes simultaneously.  The four CHARA beams are filtered by
single-mode fibers and the beams are rearranged into a 1-dimensional
non-redundant pattern and brought to a focus.  These overlapping beams
create six interference fringes, each with a unique spatial frequency.
The pattern is then focused by a cylindrical lens 
 into a ``line'' of fringes which are subsequently dispersed
by a simple spectrograph with spectral resolution $\Delta\lambda\sim0.035\mu$m.
%(spectral resolution $R=\lambda / \Delta\lambda \sim 44$).
Fast readout of the Rockwell PICNIC camera (frame time 5.5~ms)
effectively freezes the atmosphere under most seeing conditions in the
infrared.  In this way, MIRC can measure 6 visibilities, 4 closure phases, and
4 triple amplitudes simultaneously over 8 spectral channels spanning the
astronomical H-band ($\lambda=1.50-1.74\mu$m).
%The AMBER combiner on the Very Large Telescope Interferometer
%(VLTI)
%shares the same basic beam combiner architecture \cite{petrov2007} as MIRC.

Here we briefly outline the MIRC data analysis method.  After background
subtraction, the fringe patterns are analyzed by taking the Fourier
transform.  From this intermediate data, the fringe phases and
amplitudes can be combined to form the triple product, often
expressed as complex number
that can be coherently averaged (the angle argument is the closure
phase)\cite{monnier2003}. The power spectra are also accumulated for visibility-squared
estimation.  
%In the current pipeline, we do not explicitly make use of
%differential quantities or closure amplitudes.  
Bias in the power
spectrum is subtracted using a combination of ``foreground'' observations (data
taken with halted delay lines) and measurements using high spatial
frequencies immune to contamination by true fringe power.

The above procedures result in tabulations of uncalibrated
squared-visibilities and (complex) triple amplitudes.  In order to
calibrate the amplitudes of these quantities, we must estimate how much
light is injected into the fibers during fringe measurements.  We use 
spinning choppers to
partially obscure each input pupil during fringe acquisition,
chopping each beam at a unique 
frequency (25 Hz, 30 Hz, 35 Hz, 40 Hz).  Since fringes are
spatially-modulated, we can use the temporally-chopped
intensities  to obtain an estimate of the fiber coupling
efficiencies simultaneous with fringe measurements.

At this stage in the analysis, individual data files have been calibrated
but no estimates of the system visibilities have been made. Since
we have a single-mode fiber system, the system visibilities are
highly stable, however the image-plane combiner is susceptible to
temporal decoherence since the 5.5~ms exposure time is not short enough to
completely freeze turbulence.  
We track these  and any other changes in system visibility 
by observing calibrator
objects with known sizes, in this case $\gamma$~Lyr and $\upsilon$~Peg with
estimated uniform disk diameters of $0.74\pm0.10$~mas \cite{leggett1986} and
$1.05\pm0.05$~mas (MIRC/CHARA)
respectively.  While uncertainties in calibrator diameters are 
often the dominant error for 
long baseline interferometers like the CHARA Array, such errors are generally not
important for Altair which is unusually highly-resolved 
-- our dominant errors are from
imprecise calibration of mean fiber coupling and fast changes in 
atmospheric coherence time.

%While our absolute precision of visibilities
%is limited to the 5-15\% level, this limitation is not important for
%low visibility data on the well-resolved object Altair.
In order to validate our pipeline and imaging procedures, we 
carried out observations of the binary $\iota$~Peg.  This binary is
well-suited for calibration, having been observed by IOTA, NPOI, PTI,
and MIRC/CHARA.  Using $\iota$~Peg we have calibrated our ``closure phase
sign'' which removes the 180\degree ambiguity in imaging.
Furthermore, we have confirmed our wavelength
calibration at the 0.3\% level through comparison with previously published 
and new PTI
measurements
\cite{boden1999}.
The top panels of Figure~S1 show the snapshot Fourier coverage
of our $\iota$~Peg observations from UT2006Sep02 as well as our calibrated
visibility-squared data compared to the binary model fit.

$\iota$~Peg represented a suitable target for validation of the MACIM 
imaging algorithm \cite{macim2006} as
well.  The bottom panels of Figure~S1 show the MACIM image reconstruction 
of the $\iota$~Peg binary along with
the best-fit binary model; these results are consistent with the prediction
from the orbit \cite{boden1999}.
The MACIM image is in excellent agreement with the model,
including the size determinations of the two stars.  We also show the
imaging results using the CLEAN algorithm \cite{hogbom1974} with self-calibration.  A detailed
analysis of $\iota$~Peg will follow in a subsequent paper.

% vmags 0.765 (2% err)
% hmags 0.235 (lit) 0.221 (fix), 0.216 (free)    4\% err

Since imaging with optical interferometry is still novel, we present here
the calculated interferometric observables from our MACIM/MEM image
presented in Figure~2 of the mainn report. Figures S2-4 contain all the individual data
points used in this Report and they are compared with the imaging
results. 
As found in other studies \cite{iota2004,mirc2006}, the closure
phase quantities are particularly robust and precise,
showing none of the calibration difficulties typically encountered
for measurements of visibility amplitudes.
%no obvious signs of calibration difficulties.  
The calibrated data for Altair, stored in the
Optical Interferometry data exchange format (OI-FITS) \cite{oifits2005}, 
are available from the authors.

The final topic to discuss is the special imaging procedure for
Altair. Firstly, we
emphasize that use of MEM for interferometric imaging is standard
practice \cite{mem1986,tuthill2000} and the specific program MACIM
\cite{macim2006} has been validated on other test data
\cite{beauty2006}.  Hence, we will not explain the fundamentals of
interferometric imaging here or why specialized software is required for
optical interferometers \cite{oifits2005}, but rather refer the reader to the
extensive literature \cite{cornwell1981,pearson1984,baldwin1996,hummel2003,iota2004}.

For imaging, we treated each wavelength channel as providing a distinct set of
(u,v) plane coverage, ignoring any wavelength-dependence of the image itself -- this
procedure is sometimes referred to as wavelength-super-synthesis.
This assumption is well-justified for infrared intensities of hot stars
since the relative
intensities across the photosphere for the Altair model are nearly identical at 
1.5$\mu$m, 1.65$\mu$m, and 1.8$\mu$m, showing relative distortions of $<$0.5\%.
This level of inaccuraciy is much less than our observed temperature reconstruction
errors of 4\%.  Note that our von Zeipel 
modeling code did treat this wavelength-dependence better by splitting the
H band into 4-different sub-bands for fitting to the wavelength-dependent visibility
and closure phases data.  

The main difficulty in imaging the surface of a star is that the
photospheric emission is expected to show a sharp fall-off at the
limb.  In terms of Fourier modes, this sharp cutoff is encoded in very
long baseline visibilities which can not be observed.  This in
combination with the MEM procedure causes extensive ``spreading out''
in a reconstructed image, with more spreading happening where we
lack the longest baseline data.  From this perspective, we identify
contradictory goals for the imaging procedure -- smooth out the image
as much as possible except right at the edge where we expect the
sharp cutoff in emission.
This problem is similar to that encountered 
by others \cite{monnier2004} attempting to image diffuse  circumstellar
material surrounding an unresolved point source.  In the latter case, the
imaging procedure was stabilized by using a point-source model as an
``image prior'' to the MEM procedure, based on a priori
knowledge of the target under scrutiny. 
% reference to asteroid deconvolution?

For the imaging reported here, we used a uniform ellipse as a prior to the
MACIM/MEM imaging.  For the given elliptical prior, we ran the MACIM/MEM
algorithm and found the image with maximum entropy fitting the data with a 
$\chi^2_\nu\sim1$.   This procedure was robust --  the MEM prior
naturally limited the flux inside the elliptical boundary while the Maximum Entropy
maximization tended to spread out the flux as much as possible consistent with the
data itself.

% \begin{figure}[ht]
% \begin{center}
% \includegraphics[angle=0,width=6in,clip]{Figures/science_supps_im.epsi}
% \caption{\footnotesize Maximum Entropy of final MACIM/MEM image reconstruction
% as a function of the major and minor axis of the uniform elliptical prior adopted.
%There is a well defined
% }
% \End{center}
% \end{figure}

The main complication in applying the above procedure is that we
do not know {\em a priori} which ellipse to choose for our MEM prior.  One
could use the uniform ellipse derived from short baseline data, e.g.,
from the PTI data of Van Belle et al. \cite{vanbelle2001}; however,
one realizes that this is not optimum since the best-fit uniform
ellipse underestimates the actually photospheric boundary of an oblate
star with gravity darkening (in the case of Altair by $\sim$5\%).  In
order to keep our imaging procedure general and avoid a bias through our
choice of one specific elliptical prior, we carried out MACIM/MEM
imaging on a grid of 500 different uniform ellipses spanning a range of possible
sizes, elongations, and position angles.  As expected, the
``entropy'' of the final image varied depending on the prior we
adopted and it was a simple matter to find the global Maximum Entropy image
from the ensemble. 
%Figure~8 shows a summary of this grid search near the location of our final MEM image.   
%Note that the best choice for the prior
%was straightforward and robust to determine, aided by our long baselines that sample the visibility null in multiple directions.

As a final check on our calibration consistency, we carried out the above
imaging  procedures
on the MIRC/CHARA data split by observing night.  Figure~S5 shows the final MACIM/MEM images
for the two nights separately.
Based on the variation between the two independent images, we estimate
the photometric uncertainty in the final reconstruction to correspond to
$\pm$4\% in intensity across the photosphere (with a worst case
$\pm$10\% -- near the limb of the star).
The high degree of similarity gives us confidence that the final
image reconstruction is not corrupted by 
night-to-night calibration errors.   

Lastly, we comment on some confusion in the literature.  Unfortunately
the first published results on Altair \cite{vanbelle2001}
inadvertently had the (u,v) coordinates switched.  This mistake was
compounded in the next paper on Altair, Ohishi et
al. \cite{ohishi2004} from NPOI, which also made a coordinate mistake.
These errors were noticed by Domiciano de Souza \cite{souza2005} who
attempted a correction in order to combine all the data together in a
self-consistent way (although this was not mentioned in the paper
itself).  Most recently, Peterson et al. \cite{peterson2006b}
re-analyzed the original NPOI data, correcting the UV coordinate
mistakes and pointing out the original PTI errors.  However, this
paper appears to have gotten the closure phase calibration incorrect
-- causing a 180\degree rotation in their published synthetic model images.
In most cases, these errors affected only the inferred viewing
orientation of Altair, thus they did not impact the astrophysical
interpretation of the Altair data.

% Following is a new environment, {scilastnote}, that's defined in the
% preamble and that allows authors to add a reference at the end of the
% list that's not signaled in the text; such references are used in
% *Science* for acknowledgments of funding, help, etc.

% For your review copy (i.e., the file you initially send in for
% evaluation), you can use the {figure} environment and the
% \includegraphics command to stream your figures into the text, placing
% all figures at the end.  For the final, revised manuscript for
% acceptance and production, however, PostScript or other graphics
% should not be streamed into your compliled file.  Instead, set
% captions as simple paragraphs (with a \noindent tag), setting them
% off from the rest of the text with a \clearpage as shown  below, and
% submit figures as separate files according to the Art Department's
% instructions.

% \clearpage

% \noindent {\bf Fig. 1.} Please do not use figure environments to set
% up your figures in the final (post-peer-review) draft, do not include graphics in your
% source code, and do not cite figures in the text using \LaTeX\
% \verb+\ref+ commands.  Instead, simply refer to the 
%figure numbers in
% the text per {\it Science\/} style, and include the list of captions at
% the end of the document, coded as ordinary paragraphs as shown in the
% \texttt{scifile.tex} template file.  Your actual figure files should
% be submitted separately.

\clearpage

\begin{figure}
\begin{center}
\includegraphics[angle=90,width=5in,clip]{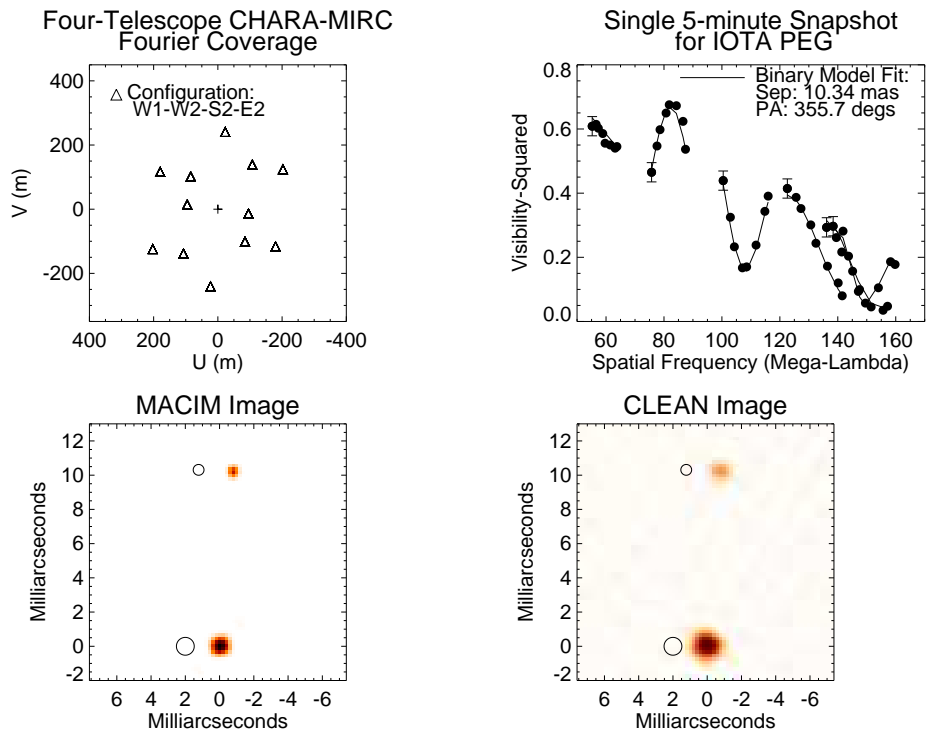}
\caption{\footnotesize These four panels validate the MIRC/CHARA pipeline and the MACIM
image reconstruction software. The top left panel shows the snapshot 4-telescope  Fourier coverage for
an observation of the calibration binary $\iota$~Peg on UT 2006 Sep 02.  The top right panel shows
the calibrated squared-visibility data along with the best-fit binary model (representative errors are shown
only for long-wavelength channel for clarity).  The bottom panels show a comparison of the image reconstructions using the MACIM and CLEAN algorithms with the best-fit binary model (circles offset 2 mas to the east).\label{fig4}}
\end{center}
\end{figure}

\clearpage

\begin{figure}
\begin{center}
\includegraphics[angle=0,width=5in]{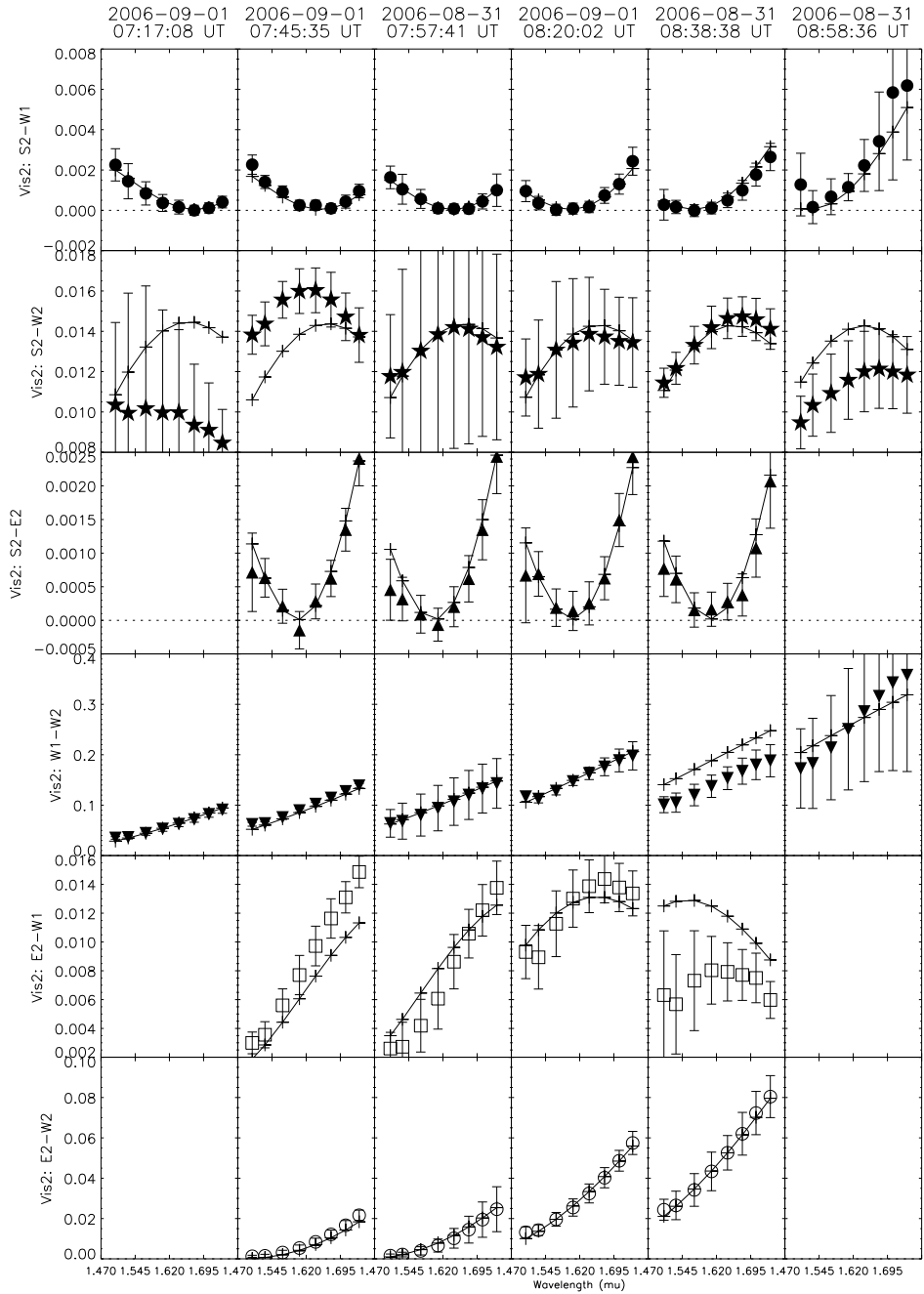}
\caption{\footnotesize This figure shows the squared-visibilities (with errors)
observed for Altair along with the calculated values from the MACIM/MEM image
(line with crosses)
presented in left panel of Figure~2.  Each column is a different observing time
while each row represents a different baseline.  Inside each panel, the
x-axis shows the wavelength of the spectrometer channels.
Note that the visibility nulls shown above for baselines S2-W1 and S2-E2
are {\em second nulls} while the visibilities in E2-W2 and E2-W1 baselines are slightly
before the first null.
}
\end{center}
\end{figure}

\clearpage

\begin{figure}
\begin{center}
\includegraphics[angle=0,width=5in]{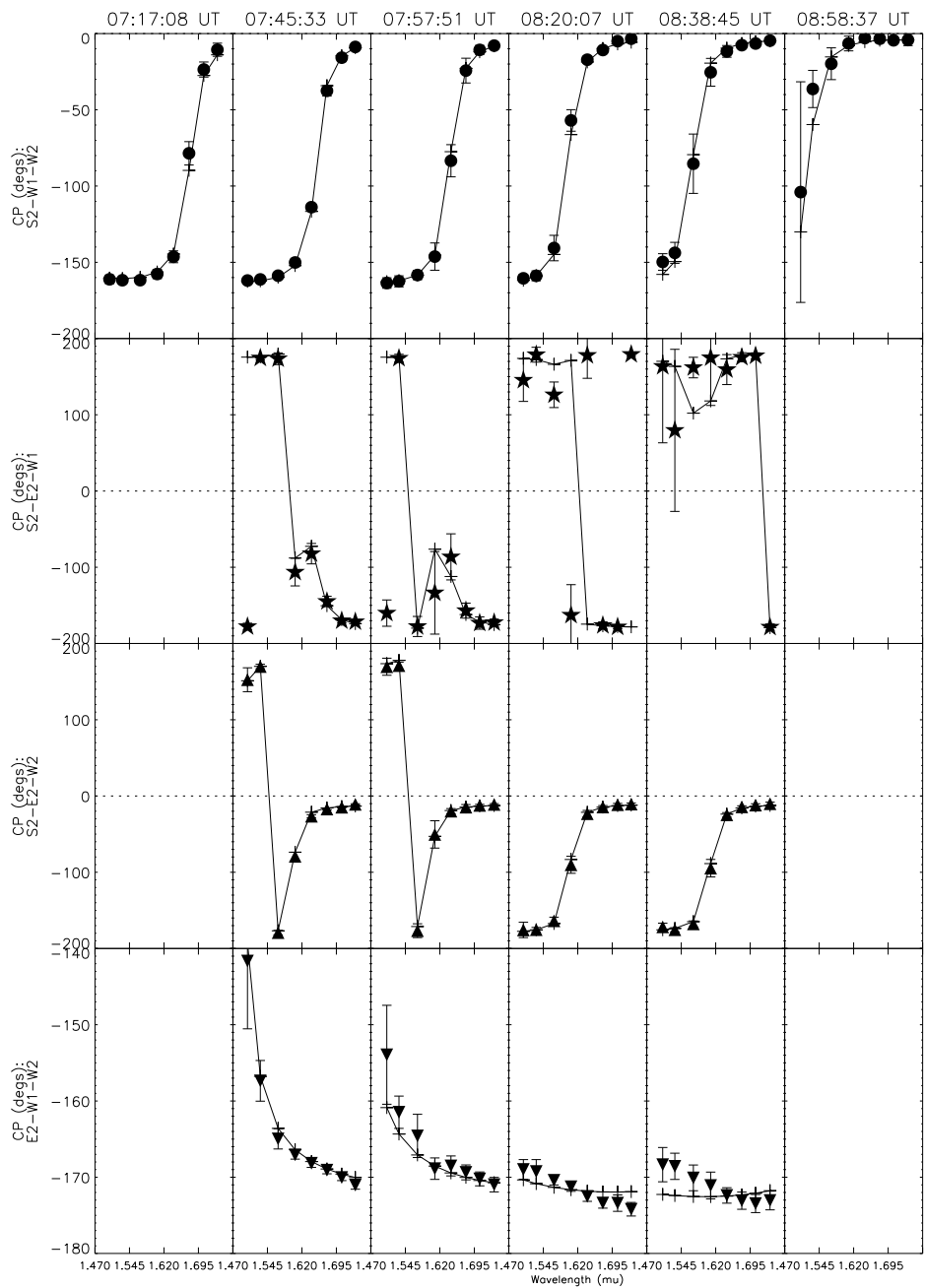}
\caption{\footnotesize All closure phase measurements
are shown for the Altair observations along with results from MACIM/MEM image (line with crosses).  Note that the closure phase has a 360\degree phase ambiguity,
thus a phase of +180\degree and -180\degree are identical in the panels above.
The columns are different times and the rows represent different closure triangles.
}
\end{center}
\end{figure}

\clearpage

\begin{figure}
\begin{center}
\includegraphics[angle=0,width=5in]{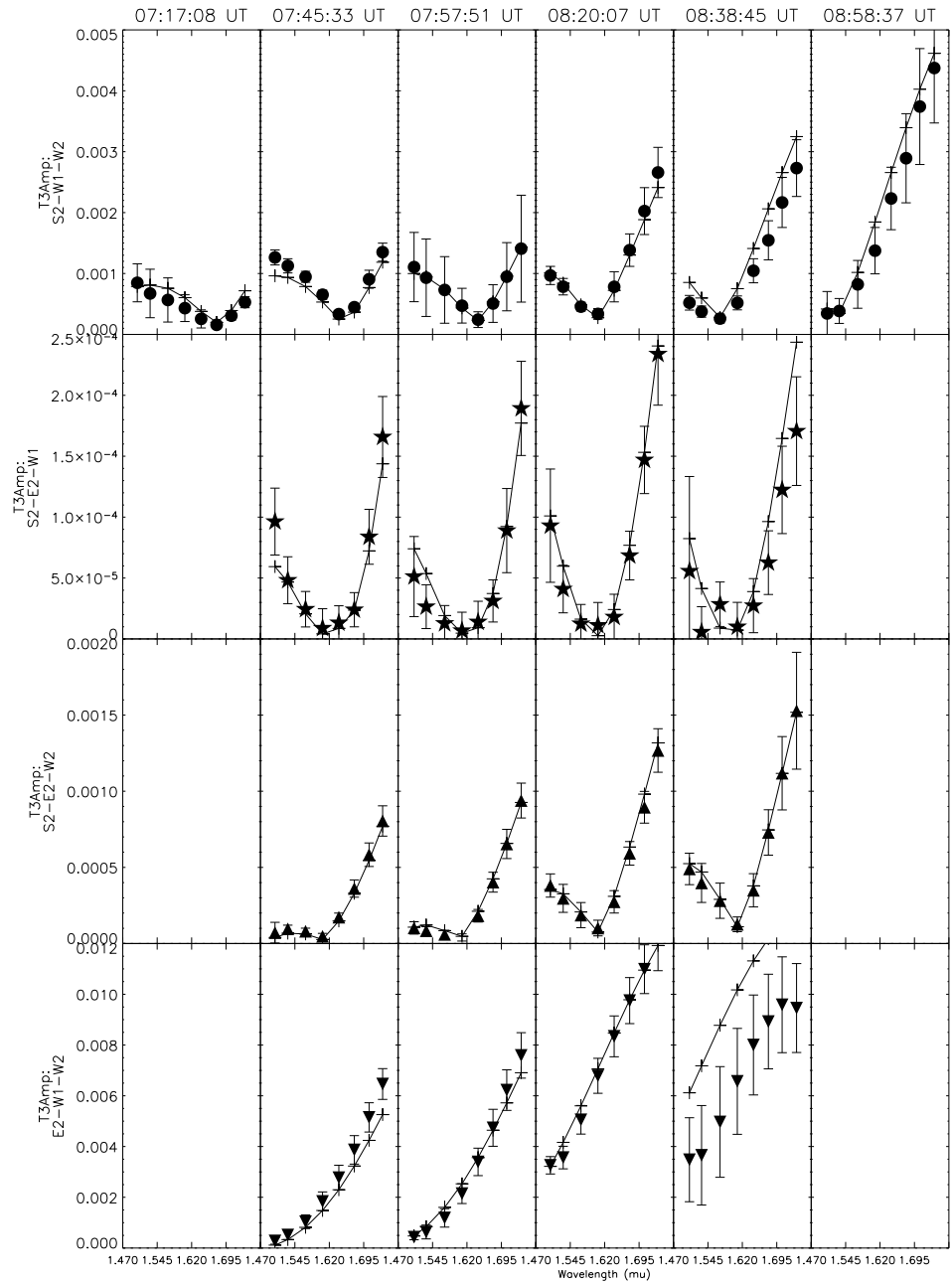}
\caption{\footnotesize All triple amplitude measurements
are shown for the Altair observations along with results from MACIM/MEM image (line with crosses).
The columns are different times and the rows represent different closure triangles.
}
\end{center}
\end{figure}

\clearpage

\begin{figure}
\begin{center}
\includegraphics[angle=90,width=6in]{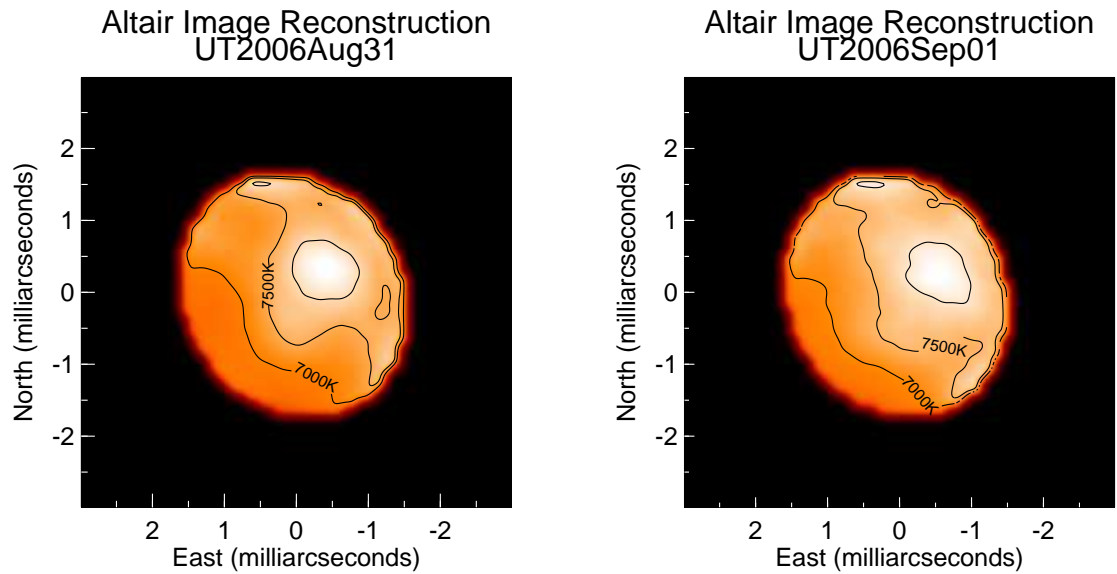}
\caption{\footnotesize Similar to Figure 2 of main Report,
except here we compare
imaging results from two independent data sets.  These image reconstructions
agree at the 4\% level rms, with maximum deviations of 10\% near the limb.
}
\end{center}
\end{figure}

\hphantom{\cite{ackno}}

\bibliography{altair2007}

\bibliographystyle{Science}

\end{document}